# Issues of Regional Development and Evaluation Problems

Olexandr Polishchuk, Dmytro Polishchuk, Maria Tyutyunnyk, Mykhailo Yadzhak

Department of Nonlinear Mathematical Analysis, Pidstryhach Institute for Applied Problems of Mechanics and Mathematics, National Academy of Sciences of Ukraine, Lviv, Ukraine

**Key words:** regional development, complex system, network, evaluation, forecasting, aggregation, continuous monitoring, parallel algorithm

## Abstract

Region of the country is treated as an open megasystem that consists of many interconnected complex network systems. Classification of such systems on the basis of ordered motion of flows is performed. The main approaches to investigation of complex systems are described. Theory of evaluation is considered as a tool for analysis of network systems. Purpose and subject of evaluation of the systems are formulated. The main problems of evaluation are analyzed and the ways of further research are determined.

## 1. Introduction

Fostering the development of depressed regions of the country or recovery of regions, that suffered from climatic, geological or technological catastrophes as well as local military conflicts require involvement of substantial financial, material and human resources [1, 2]. Each of those regions represents an example of open megasystem that is a combination of many interconnected systems of various types and purposes functioning with relation to each other. The systems considered include economic, infrastructural, transport, social, informational, etc [3-6]. Most of them have network structure and hierarchical control system [7, 8]. The role of every system in life of the region may vary. In some cases, improvement of state and operational quality of only one of the systems (e.g. transport) may pose substantial impact on the development of the whole region [9]. More complicated situation arises when it is necessary to improve most of the megasystem components. Limited resources do not allow to perform actions for their development in all areas simultaneously and with equal intensity. Thus, the problem arises consisting in determination of priority of systems and their components during planning and organization of operational sequence, direction of the financial flows required, etc.

Analysis of state and operational quality of megasystems, such as regions of the country, is a complex problem. Systems comprising the megasystem usually are of different type, purpose, composition and structure. They function according to different rules and require usage of different research methods. Among many disciplines involved in the study of complex systems (systems analysis, systems theory, theory of complex networks, decision making theory, mathematical modeling methods [10-13], etc.), the theory of evaluation holds a special place [14]. It reached its most rapid development during last decades: existing systems expand, become more complex and form megasystems, whose state and operational quality depend on the quality of life and security of citizens. We understand evaluation as quantitative expression of results of multi-criteria analysis of the state, operational quality and interaction of the elements of complex systems. Such evaluation is an objective basis for making informed decisions about further actions regarding system or its components [12].

In this article we try to formalize the procedure for evaluation of network systems that comprise the megasystem of region and to describe main directions, results and prospects of our research.

## 2. Complex Networks and Network Systems

In the recent years, the theory of complex networks is developing rapidly [11]. The term "network" usually stands for the set of nodes connected by some relations or edges (if graphic representation is available). Complex network is often called "the system". However, let us imagine a railway network as a set of stations and railway tracks connecting them where the railway traffic is absent or a gas or oil supply network as a set of compressor stations and pipelines connecting them which are not pumped with gas or oil. How would computer network represented as a set of servers, computers and wire or wireless connection means look like, if it wasn't used for exchange of information? Lot of further examples may be provided. Presence of material and/or information flows is one of the features that characterize certain object as a system [13]. The purpose of creation, operation and development of any network is to provide the motion of flows, i.e. it is the motion of flows that makes complex network a system. In this case, network is only a "frame" of the system. The composition and structure of the regional networks in the process of their development are dynamic with regard to changing the number of nodes and the relations between them.

All network systems (NWS) regardless of their type and purpose can be differentiated according to flow organization. We distinguish systems with three levels of organization: fully ordered motion of flows, partially ordered motion of flows and disordered motion of flows. Railway transport system of country and power supply systems are examples of NWS with fully ordered motion of flows. Road transport system of large city or country, postal and trade networks are examples of NWS with partially ordered motion of flows. Social networks, electronic media, mobile communication systems, etc. are examples of NWS with disordered motion of flows. Complex networks with the same structure and different levels of ordered motion of flows generate different network systems.

Organization of flows must be maintained. This means that control of state, operation quality and interaction between objects that comprise the system and provide motion of flows in the network is required. Such functions are performed by control system of NWS. Control system (CS) can be organized according to territorial, operational or hybrid principle. CS and NWS controlled by it together form hierarchical network system (HNWS). Most of industrial, transportation, financial and other systems created and controlled by human kind are the HNWSs. Specific feature of HNWS is that each subsystem of some hierarchal level is divided into a set of subsystems that create subnetwork of the network of lower level [15]. At each level of the hierarchy, the edges ensure smooth motion of flows of certain type, whereas the nodes ensure their processing. The lowest level of hierarchy provides motion of flows for which the network was created (trains, cars, energy resources, information etc.). Information, organizational and administrative decisions, cash flows, etc. are flows of higher levels of hierarchy (control levels). If we return to the example of the region as megasystem, we may note that different NWS creating it have different composition, structure and level of ordered motion of flows.

Different ordering of motion of flows requires usage of different approaches to the analysis of network system operation process. These approaches can be based on deterministic, statistical, stochastic and hybrid principles [16-19]. Each of them has its own advantages and disadvantages [20]. Purpose of deterministic methods consists in formulation of conclusions regarding actual state and operational quality of each system element (node, edge, flow) [21, 22]. It is often impossible to perform detailed analysis of all objects of the system. For example, comprehensive medical examination of all citizens of the country is very difficult to carry out. It is also impossible to track the traffic of all vehicles in megalopolis. At the same time, such researches are extremely important for planning production and procurement of medications and medical devices or for improvement of efficiency of transport system in large cities. In such cases, the statistical methods are used [23]. The validity of statistical studies of general set depends on thoroughness of the deterministic analysis of objects included in the representative set If information about system is incomplete or unclear, the stochastic methods are applied. In this case, preliminary data on the distribution of probabilities are usually obtained from results of statistical studies. The history of nuclear power started over half the century ago. At the beginning of its development, it was estimated that the probability of a serious nuclear reactor accident comprises 1 accident in 10 million years [24]. During the operation of nuclear reactors there were at least 15 serious accidents and catastrophes. The number of nuclear reactors operating throughout the word reached 438, and their total operational time is by times lower than 10 million years. It should also be taken into account that there are many systems that require usage of solely deterministic research methods. Indeed, using probabilistic approach to diagnostics or airliner onboard systems check doesn't seem appropriate. Deterministic methods of research are used for analysis of all NWSs, regardless of the level of ordered motion of flows, however the scope and purpose of these studies are different.

## 3. Purpose and Subject of Evaluation of Complex Network Systems

Usually, the most popular purpose of evaluation consists in the search of system objects whose operation is unsatisfactory [20]. The notion "system object" will hereinafter designate structural unit of system of any hierarchy level – from element to the subsystem of highest level of splitting. These objects have negative impact on all related system components. Operation of HNWS may be significantly improved through improvement of such objects. Search of system objects that operate perfectly is also important. Those objects may be then used as references. Expanding of principles of their operation organization to other similar objects of HNWS also contributes to improvement of its general operation. Novelty detection is the purpose of the study of some systems. This means for example search for atypical objects on the basis of satellite, GPR or aerial images. Most of the systems created and controlled by human kind can operate in different modes. The purpose of evaluation may consist in determination of the most appropriate or extreme (critical, dangerous) modes. The same mode may be both appropriate and critical for various state of the network. High-speed train movement does not pose any danger to railway of high quality and can cause a catastrophe if the state of track is unsatisfactory. Purpose of evaluation may also consist in choosing the best system from a given class of equivalent systems [20].

The subject of evaluation of each system object of any hierarchy level is its
- state;
- operational quality;
- efficiency of interaction with other objects of system.

These signs are interrelated and mutually dependent. Indeed, it is difficult to expect high-quality operation of object, if its state is unsatisfactory. Objects that operate unsatisfactory have negative impact on all system elements interacting with them. Combination of the results of evaluation of the state and operational quality of object as well as its interaction with other components provides a fairly complete and integral idea of object operational quality [25].

### 4. Main Evaluation Problems for Complex Network Systems

Let us list main problems that involve evaluation of complex systems:

1. *Evaluation of state and operational quality of system element.* Solution for this problem allows to determine elements representing potential threat for operation of system in general and those capable of causing failures as well as to analyse their impact on surrounding elements [26, 27]. If system is consists of elements of the same type, solution of this problem allows to determine elements operating in the best way, i.e. reference elements. Finally, development of generalized conclusions regarding general system operation quality is based on results of evaluation of system elements.

2. *Forecasting state and operational quality of system elements.* State and operational quality of element usually changes in course of time. These attributes can cross "safety threshold" and pose danger to separate components of the system. Forecasting the behavior of the evaluations may be short- and long-term. In any case forecasting term must provide possibility to correct possible faults.

3. *Choice of optimal mode for system operation.* Solution of this problem allows to determine both most appropriate and extreme system operating modes, as well as modes of potential failure [28].

4. *Evaluation of system state and operational quality.* Solution of this problem allows to determine general quality of system operation according to defined set of parameters, criteria and operating modes [25].

5. *Choice of optimally operating system from given class of equivalent systems.* Solution of this problem allows to determine the best (reference) or the worst systems in class. Optimally operating elements, modes and systems determined during evaluation may be used as practically reachable quality references [29].

6. *Analysis of system operation history.* Solution of this problem allows to track and forecast the quality of system operation, to determine trends of its development in the context of improvement or deterioration and to prevent possible failures [30].

List of evaluation problems for each particular system can be expanded with regard to its features and purpose of the research.

### 5. Methods for Evaluation of Complex Network Systems

Usually, two main approaches are applied to control state and behavior of existing regional HNWS: regular scheduled inspections, distinctive features of which are accuracy and possibility for further development of recommendations for elimination of drawbacks discovered; and continuous monitoring of system objects' functioning that allows us to draw mediate, but still significant conclusions regarding its actual state and functioning quality [31].

It is reasonable to start evaluation of real systems with objects of lowest structural level, i.e. with elements of HNWSs. We define an element as an object of clearly defined location, functional destination and relevant set of characteristics describing its state and functioning process with corresponding ranges of permissible values for those characteristics. All characteristics are evaluated according to certain collection of criteria and parameters. Of course, evaluation of every object presupposes evaluation of its state on the first place, and only after that the evaluation of quality of implementation of its functions that in any case depend on element's state – either directly or indirectly. The process of evaluation is started only after the stage of thorough selection and processing of experimental data as to each of characteristic and their conversation into format, suitable for further analysis.

Currently, for evaluation of HNWS integer rating or conceptual ("excellent", "good", "satisfactory", "unsatisfactory") scale [14] is commonly used. Its main drawback is that "satisfactory" evaluation may imply wide range of concepts – from "almost good" to "slightly better than unsatisfactory". We propose [30] unified approach for evaluating state, quality of functioning and interaction between system structural elements, which consists in developing main rating evaluation and its adjustment with regard to type and features of object studied. Such an approach allows not only to compose more clear understanding of evaluated object, but also to localize the reasons for drawbacks discovered.

The number of characteristics describing element may comprise dozens [20]. Different characteristic may be selected in different ways and they priority regarding structure and functions of element may be different. It is clear that the conclusions as to separate characteristics are to be generalized with consideration of their priority. Recording the number of actually evaluated elements' characteristics is also important. From this point on, evaluations for elements' state and functions they implement on the basis of their characteristics behavior analysis will be referred to as *local* [32].

As usual, scheduled inspections of system's objects are held at different time points, which means the results of last study may not stay on such stage till following inspection, and state of object and its functioning quality may cross "safety threshold". It should be also taken into account that every real system evolves in time, i.e. with regard to current requirements, its evaluation may be insufficient. Therefore, evaluation process should contain means of analysis of systems meeting expected requirements for short- and long-term perspective. Thus, the evaluation process should not only determine conclusions and discover "faulty" elements for the time point moment when study is held, but also it should forecast further behavior of system objects. *Forecasting* analysis performed on the basis of local evaluations prehistory, allows us to determine the nature, direction and speed of system state change, follow up negative processes and forecast potential risks, as well as material and financial expenses required for their elimination or timely prevention [32].

Number of local evaluations of real HNWS may reach dozens of millions values [20], which obviously exceed the capacity of their manual analysis. For their generalization, i. e. for developing conclusions regarding their state, quality of functioning and interaction of objects of higher hierarchy levels (subsystems and HNWS in general), tools of linear and non-linear aggregation are applied [30], taking into account weighted coefficients that reflect importance of separate objects in system's structure and priority of functions they perform. Since weighted averaging mitigates the results of both positive and negative evaluations, it is reasonable to make generalization of conclusions after elimination of causes and revaluation of drawbacks eliminated. Let us refer to above described method as to *aggregated* [33].

Due to the number of reasons, scheduled inspections may often not discover drawbacks that arise "out of schedule". It should be also taken into account that even excellent state and functioning quality of separate objects in the system do not ensure high performance of its subsystems or system in general. And vice versa, the most optimal work organization process will not ensure high efficiency of system functioning if HNWS's state or organization of components functioning is unsatisfactory. The more worn-out HNWS's objects are the more urgent is the problem of continuous monitoring of their state and functioning process. Quality of implementation of functions by object may be affected by number of third-party factors, both internal and external as to the system. Internal influence may be evaluated on the level of subsystems connecting interacting objects. We shall call this evaluation method *interactive* [34]. It allows us to determine separate objects in selected subsystem, functioning of which is unsatisfactory, without thorough analysis of state and functioning quality of these objects and expenses related to such analysis. The simplest interactive evaluation may be performed for system where the movement of flows is deterministic, at least partially, in accordance with certain schedule, the compliance to which may be periodically summed up. It is reasonable to include generalized results of interactive evaluation over certain time period between two scheduled inspections into aggregated evaluation procedure. Those results may be also used for more detailed and accurate forecasting analysis of functioning of evaluated system's objects.

In general, only if combined, proposed methods may provide sufficiently full and adequate understanding of HNWS quality. Indeed, high local evaluations do not ensure effective interaction of elements, failures of separate systems objects may result in breakdown in balanced organization, satisfactory state of object for the moment of current inspection does not imply the state will stay satisfactory till the next inspection. Huge amount of information regarding separate HNWS elements without appropriate generalization is ill-suited for rapid analysis and timely reaction for drawbacks discovered. On higher generalization levels, evaluation allows to determine reliable conclusion as to the state and functioning quality of system and its main subsystems and to define measures, as well as material and finance expenses required for its modernization and optimization of functioning. At the local level evaluation allows to identify separate elements and their components subject to improvement.

While processing huge amounts of data that are to be analyzed and evaluated on the real time basis the problem of calculation optimization arises. This problem is solved through parallelization of algorithms of system evaluation. In [35], we proposed and studied parallel-sequential approach for calculation optimization (in time) during local evaluation of quality of HNWS operation. This approach is intended for implementation on parallel computer systems with shared memory. In [36], algorithmic constructions were developed for aggregative evaluation of behavior of elements, separate subsystems and whole systems. These constructions are intended for implementation on parallel computer systems with shared and distributed memory. The proposed approaches to organization of parallel calculations take into account actual capabilities of computer system (number of nodes and node cores, node memory capacity, communication network performance, etc.).

## 6. Future Research

At the moment, authors have developed methodology for complex deterministic evaluation of regional HNWSs with fully ordered motion of flows. The next stage of our work consists in development of methodology for evaluation of systems with partially ordered and disordered motion of flows. The less ordered is the motion of flows in the system and the more dynamic is its structure, the more difficult is the problem of studying this HNWS and the more indefinite are the results of study. Some systems require development of specific evaluation methods that combine deterministic, statistical, and stochastic approaches.

Today computer networks with integrated powerful parallel multi-purpose tools (in most cases these are cluster systems) are being created and developed. Clusters are the most effective way to implement parallel algorithms as a sets of independent or loosely coupled branches. Today we work on development of above mentioned algorithms [37]. This would help to optimize time consumption during application of methods for state and operation quality evaluation and forecasting in actual regional HNWSs.